\documentclass[aps,pra,twocolumn,showpacs,preprintnumbers,superscriptaddress]{revtex4} 
\usepackage{graphicx,bm,dcolumn,amsmath,amssymb,amsfonts} 
\newcommand{\vm}{\bm{m}}

\newcommand{\vsigma}{\bm{\sigma}}

\begin{document} 
 
\title{Geometric quantum gates in liquid-state NMR based on a cancellation of dynamical phases} 
 
\author{Yukihiro Ota} 
\altaffiliation[Present address: ]{CCSE, Japan Atomic Energy Agency, 
6-9-3 Higashi-Ueno, Tokyo 110-0015, Japan and 
CREST(JST), 4-1-8 Honcho, Kawaguchi, Saitama, 332-0012, Japan}
\affiliation{ 
Research Center for Quantum Computing, Interdisciplinary Graduate School  
of Science and Engineering, Kinki University,  
3-4-1 Kowakae, Higashi-Osaka, 577-8502, Japan} 
\author{Yoshito Goto} 
\affiliation{ 
Interdisciplinary Graduate School  
of Science and Engineering, Kinki University,  
3-4-1 Kowakae, Higashi-Osaka, 577-8502, Japan} 
\author{Yasushi Kondo} 
\affiliation{ 
Research Center for Quantum Computing, Interdisciplinary Graduate School  
of Science and Engineering, Kinki University,  
3-4-1 Kowakae, Higashi-Osaka, 577-8502, Japan} 
\affiliation{ 
Department of Physics, Kinki University, 
3-4-1 Kowakae, Higashi-Osaka, 577-8502, Japan} 
\author{Mikio Nakahara} 
\affiliation{ 
Research Center for Quantum Computing, Interdisciplinary Graduate School  
of Science and Engineering, Kinki University,  
3-4-1 Kowakae, Higashi-Osaka, 577-8502, Japan} 
\affiliation{ 
Department of Physics, Kinki University,  
3-4-1 Kowakae, Higashi-Osaka, 577-8502, Japan} 
 
\date{\today} 
 
\begin{abstract} 
A proposal for applying nonadiabatic geometric phases to quantum
 computing, called double-loop method [S.-L. Zhu and Z. D. Wang,
 Phys. Rev. A {\bf 67}, 022319 (2003)], is demonstrated 
in a liquid state nuclear magnetic resonance quantum computer. 
Using a spin-echo technique, the original method is modified so that
 quantum gates are implemented in a standard high-precision nuclear
 magnetic resonance system for chemical analysis. 
We show that a dynamical phase is successfully eliminated and a
 one-qubit quantum gate is realized, although the gate fidelity is not high. 
\end{abstract} 
 
\maketitle 
 
\section{Introduction} 
Precise and reliable control of a quantum system is an attractive and
challenging experimental issue in quantum physics. 
In particular, the importance of its application to quantum information
processing has been increasing recently. 
A promising way to achieve this is to employ geometric
phases (or, more generally, non-Abelian
holonomies)\,\cite{Nakahara2003,ChruscinskiJamiolkowski2004}, because 
geometric phases are expected to be robust against noise and decoherence
under a proper condition\,\cite{BlaisTremblay2003,Ota;Kondo:2009}. 
A large number of studies for applying their potential
robustness to quantum computing have been done, e,g., 
phase-shift gates with Berry phases\,\cite{jones2000}, 
nonadiabatic geometric quantum 
gates\,\cite{Wang;Keiji:2001,Zhu;Wang:2002,zhu2003,Tianetal2004,
Ding;Wang:2005,ZhuZanardi2005,Das;Kumar:2005,Imai;Moriaga:2007}, 
holonomic quantum computing\,\cite{Zanardi;Rasetti:1999,Duan;Zoller:2001,
Niskanen;Salomaa:2003,Solinas;Zanghi:2004,Karimipour;Majd:2004,
Tanimura;Hayashi:2005,Goto;Ichimura:2007,Ota;Nakahara:2008}, 
quantum gates with noncyclic geometric
phases\,\cite{Friedenauer;Sjoqvist:2003}, and so on.
 
For physical realization of geometric quantum gates, elimination
of a dynamical phase is a key point. 
Jones {\it et al.}\,\cite{jones2000} implemented a controlled phase-shift
gate with a Berry phase\,\cite{Berry,Simon:1987} 
by quasi-statistically, or slowly,  
controlling an effective field in a rotating frame. 
They nulled dynamical phase effects using a conventional  
spin echo approach\,\cite{levitt}.  
Zhu and Wang\,\cite{zhu2003} proposed a geometric quantum gate based on 
the Aharonov-Anandan phase\,\cite{AharonovAnandan1987,Page:1987},
which should be fast since a nonadiabatic process is employed. 
In their proposal, elimination of a dynamical phase is achieved 
by a double-loop method, where a dynamical phase cancels out while
a geometric phase accumulates along two loops. 

Although several experimental techniques for the application of
geometric phases to quantum computation are
available\,\cite{Tianetal2004,Das;Kumar:2005,Imai;Moriaga:2007}, explicit 
implementations of geometric phase gates have not been extensively
studied so far.  
Without explicit implementations, the often-cited advantage of the
holonomic quantum gates is nothing more than a desk plan. 
Thus, such explicit examples are highly desirable. 
In this paper, we combine Zhu and Wang's approach with Jones 
{\it et al.}'s one, employing an Aharonov-Anandan phase for fast gate operation and a
spin echo technique for dynamical phase cancellation,  
and demonstrate one-qubit gates with a commercial liquid-state  
nuclear magnetic resonance (NMR) system. 
In many experiments of nonadiabatic geometric quantum
gates\,\cite{Tianetal2004,Das;Kumar:2005,Imai;Moriaga:2007}, 
the gate operations in which the dynamical phase is arranged to
vanish\,\cite{Ota;Kondo:2009,SuterMuellerPines1988} have been adopted. 
In the present paper, we show that we may have another option for
physical realization of geometric quantum gates. 
 
The next section summarizes Zhu and Wang's theoretical proposal and  
our modifications for practical implementation in liquid-state NMR 
quantum computer. 
In Sec.\,\ref{sec:experiments} we describe details of our experiments,
where implemented gates are evaluated by performing quantum process
tomography. 
Sec.\,\ref{sec:summary} is devoted to summary. 
 
\section{Theory} 
\label{sec:theory}
\subsection{Quantum gates with orthonormal cyclic vectors} 
\label{subsec:qg_ocv}
The Aharonov-Anandan phase is a geometric phase associated with nonadiabatic  
cyclic time evolution of a quantum 
system\,\cite{AharonovAnandan1987,Page:1987}.  
Let us write a state at $t$ ($0\le t\le \tau$) as $|{\psi(t)}\rangle$ 
in the Hilbert space $\mathcal{H}$ with dimension $n$.  
The time evolution of a system is given by the Schr\"odinger equation 
with a time-dependent Hamiltonian $H(t)$. 
We take the natural unit in which $\hbar=1$. 
The nonadiabatic cyclic time evolution of the quantum system means that 
\begin{equation*}
|{\psi(\tau)}\rangle = U(\tau)|{\psi(0)}\rangle
 = e^{i\gamma}|{\psi(0)}\rangle, 
\end{equation*} 
where $U(\tau)$ is the time evolution operator and $\gamma \in \mathbb{R}$.  
Let us call $|{\psi(0)}\rangle$ a cyclic vector\,\cite{AharonovAnandan1987}. 
We note that the dynamical phase $\gamma_{\rm d}$ associated with this
time evolution is 
\begin{equation} 
\gamma_{\rm d} = - \int_0^\tau \langle{\psi(t)}|H(t)|{\psi(t)}\rangle dt, 
\label{eq:dyn_phase} 
\end{equation} 
while the geometric component is defined as 
\begin{equation*}
 \gamma_{\rm g} = \gamma - \gamma_{\rm d}. 
\end{equation*}
We take a set of orthonormal cyclic vectors, 
$\{ |{u_k(0)}\rangle \}$ so that 
\begin{equation*}
|{u_k(\tau)}\rangle =  e^{i \gamma_k} |{u_k(0)}\rangle,
\end{equation*}
where $k = 1,\ldots,n$. 
Hereafter, we write $|u_k(0)\rangle$  as $|u_k\rangle$ for brevity. 
A general state $|\Psi(0) \rangle\in\mathcal{H}$ is expressed as 
\(
 |{\Psi}(0) \rangle = \sum_{k=1}^{n}a_{k}|{u_{k}}\rangle
\). 
Then, we have 
\(
|{\Psi(\tau)}\rangle = U(\tau)|\Psi(0) \rangle= 
\sum_{k=1}^{n}a_{k}e^{i\gamma_{k}}|u_{k}\rangle
\). 
We denote a fixed basis in $\mathcal{H}$ as $\{|l\rangle \}$,
which corresponds to the computational basis $\{|0\rangle,|1\rangle\}$
for the case $n=2$. 
In terms of $\{|l\rangle\}$, we have 
\(
 |\Psi(\tau)\rangle = \sum_{l=0}^{n-1}c_{l}(\tau)|l\rangle
\). 
It means that 
\(
 a_{k} 
=
\langle u_{k}|\Psi(0)\rangle 
= 
\sum_{l=0}^{n-1}c_{l}\langle u_{k}|l\rangle
\), 
where $c_{l}=c_{l}(0)$. 
Therefore, we find that 
\begin{equation} 
c_{l}(\tau) 
= 
\sum_{l^{\prime}=0}^{n-1} 
\sum_{k=1}^{n}\,e^{i\gamma_{k}} 
\langle{l}|u_{k}\rangle \langle{u_{k}}|l^{\prime} \rangle c_{l^{\prime}} 
= 
\sum_{l^{\prime}=0}^{n-1}\,V_{ll^{\prime}}c_{l^{\prime}}, 
\label{eq:coeff_tau_cb}
\end{equation} 
where  
\begin{equation} 
V_{l  l^{\prime}} = \sum_{k=1}^{n} 
e^{i\gamma_{k}} \langle{l }|u_{k}
\rangle \langle{u_{k}}|l ^{\prime} \rangle. 
\label{eq:top_cb}
\end{equation} 
We have 
\(
  \gamma_{k} = \gamma_{k,{\rm d}} + \gamma_{k,{\rm g}} 
\), 
where $\gamma_{k,{\rm d}}$ and $\gamma_{k,{\rm g}}$ are the dynamical
and the geometric phases associated with $|u_{k}\rangle$, respectively. 

\subsection{Single-qubit case} 
Let us consider a single-qubit system.  
We choose the Hamiltonian $H_1(t)$ as  
\begin{equation*} 
H_{1}(t) = -\frac{1}{2} {\bm \Omega}(t)\cdot\vsigma, 
\end{equation*} 
with NMR in mind, where 
\(
\vsigma =
(\sigma_{x},\,\sigma_{y},\,\sigma_{z})
\) 
and
\begin{equation} 
 {\bm \Omega}(t)  =
\left( 
\omega_{1} \cos(\omega_{\rm rf} t-\phi),-
\omega_{1}\sin(\omega_{\rm rf} t- \phi), 
\omega_{0} 
\right).
\label{eq:field} 
\end{equation}
We note that ${\bm \Omega(t)}$ is a time dependent  
parameter corresponding to the external field and  
$\sigma_{\alpha}$ ($\alpha=x,\,y,\,z$) is the $\alpha$th component of the
Pauli matrices. 
One can freely control $\omega_0$ by
taking a proper rotating frame. 
The transformation to the rotating frame with the frequency
$\omega_{\rm{rf}}$ is made possible by the unitary transformation 
$U = e^{i \omega_{\rm{rf}} \sigma_z t/2}$ 
and the Hamiltonian in the rotating frame is 
\begin{equation}
H_{1{\rm r}} 
=  U^{\dagger} H_1 U - i U^{\dagger}
\frac{d}{dt} U 
= -\frac{1}{2} \Omega \vm \cdot \vsigma,
\label{eq:nmr_rot_ham}
\end{equation}
where 
\(
\Omega = (\omega_{1}^{2}+\Delta^{2})^{1/2}
\), 
\(
\Delta = \omega_{0}-\omega_{\rm rf}
\), 
\(
\vm =
( \sin\chi\cos\phi,\,\sin\chi\sin\phi,\,\cos\chi )
\), and 
\(
\tan\chi = \omega_{1}/\Delta 
\). 
The solution of the Sch\"odinger equation is 
\begin{eqnarray} 
|\psi (t)\rangle 
&=& 
e^{i\omega_{\rm rf} t\sigma_{z}/2}\, 
e^{i\Omega t \vm\cdot\vsigma/2}\, 
|\psi (0)\rangle .
\label{eq:sol_1q} 
\end{eqnarray} 
\begin{figure}[tbp]
\centering
\scalebox{0.58}[0.58]{\includegraphics{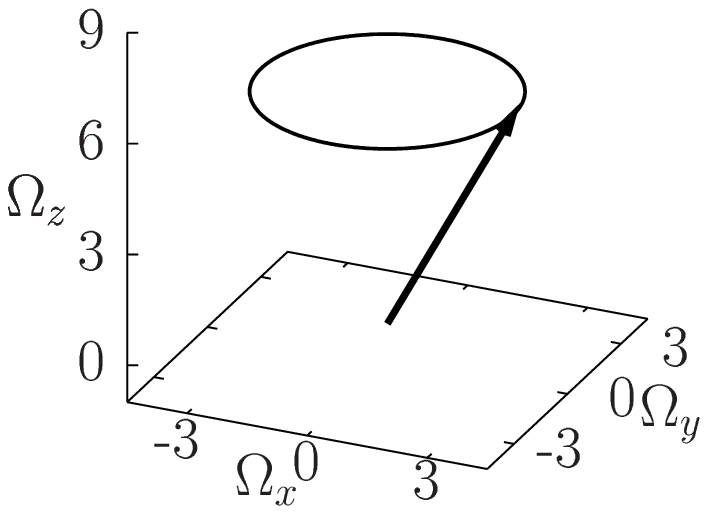}}
\scalebox{0.58}[0.58]{\includegraphics{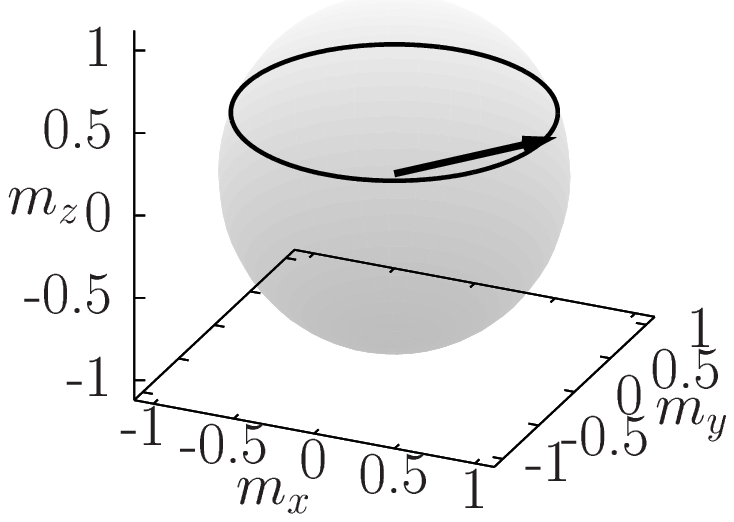}}
\vspace{-3mm}
\flushleft
(a)\hspace{40mm}(b)
\caption{Example of dynamics of a single-qubit cyclic vector. 
(a) A time-dependent external field
 $\bm{\Omega}(t)$ and (b) a closed trajectory on the Bloch
 sphere corresponding to a cyclic vector $|\psi_{+}(t)\rangle$, 
$0\le t\le \tau=2\pi/|\omega_{\rm rf}|$. 
The end point of each arrow represents the initial value. 
We set the parameters $\omega_{0}=2\pi$, $\omega_{1}=0.5\times 2\pi$,
 $\omega_{\rm rf}=0.8\times 2\pi$, and $\phi=0$ in
 Eq.\,(\ref{eq:field}). }
\label{fig:singlel}
\end{figure}
We denote the eigenstates of $\vm\cdot \vsigma$ with eigenvalues 
$\pm 1$ as 
\(
 |\psi_{\pm}\rangle
\). 
Their explicit forms are  
\begin{eqnarray*} 
 |{\psi_{+}}\rangle  
&=&  e^{-i\phi/2} \cos\frac{\chi}{2} |{0}\rangle 
 +   e^{i\phi/2}  \sin\frac{\chi}{2} |{1}\rangle ,  \\ 
 |{\psi_{-}}\rangle  
&=& - e^{-i\phi/2} \sin\frac{\chi}{2} |{0} \rangle
 +    e^{i\phi/2}  \cos\frac{\chi}{2} |{1} \rangle,  
\end{eqnarray*} 
where $|{0}\rangle$ and $|{1}\rangle$ are the eigenstates of $\sigma_z$
with eigenvalues $+1$ and $-1$, respectively, and taken as the
computational basis.  
The corresponding Bloch vectors of $|{\psi_{\pm}}\rangle$ are  
\begin{equation*} 
\langle{\psi_{\pm}}| \vsigma |{\psi_{\pm}}\rangle = \pm \vm .  
\end{equation*} 
We require that $|{\psi_{\pm}}\rangle$ be cyclic vectors. 
Since $|\psi_{+}\rangle$ and $|\psi_{-}\rangle$ are mutually orthogonal,
they are identified as $\{|u_{k}\rangle\}$ in Sec.\,\ref{subsec:qg_ocv}.  
It follows from Eq.\,(\ref{eq:sol_1q}) that the execution time $\tau$
must satisfy the condition $|\omega_{\rm rf}| \tau/2 = \pi$, i.e., 
$\tau = 2\pi/|\omega_{\rm rf}|$.
Then, $|\psi_{\pm}(\tau) \rangle$ is written as
\begin{eqnarray*} 
|{\psi_{\pm}(\tau)}\rangle = e^{i \gamma_{\pm}} |{\psi_{\pm}}\rangle, 
\end{eqnarray*} 
where 
\(
\gamma_{\pm} = \pi \pm  \pi \Omega/|\omega_{\rm rf}|
\). 
Figure \ref{fig:singlel} shows an example of $\bm{\Omega}(t)$ and the
closed trajectory on the Bloch sphere corresponding to $|\psi_{+}\rangle$. 
Using Eq.\,(\ref{eq:dyn_phase}), we find that the dynamical phase is 
\begin{equation*} 
\gamma_{\rm \pm, d}  
= \pm \frac{\tau}{2}(\omega_1 \sin \chi + \omega_0 \cos \chi)  
= \pm \frac{\pi(\omega_1^2+\omega_0 \Delta)}{|\omega_{\rm rf}| \Omega},
\end{equation*} 
while the geometric phase is 
\begin{equation*} 
\gamma_{\pm,{\rm g}}  
= 
\gamma_{\pm} -\gamma_{\pm, {\rm d}} 
=  \pi \left(1 
\mp \frac{\omega_{{\rm rf}}}{|\omega_{{\rm rf}}|}\frac{\Delta}{\Omega} 
\right).
\end{equation*} 
Based on Eqs.\,(\ref{eq:coeff_tau_cb}) and (\ref{eq:top_cb}), we obtain
a unitary gate with the dynamical and the Aharonov-Anandan phases with respect to the
computational basis $\{|0\rangle,|1\rangle\}$, 
\begin{equation*} 
V(\tau)  
=
\left( 
\begin{array}{cc} 
\cos\gamma-i\sin\gamma\cos\chi 
&  - i e^{-i \phi}\sin \gamma \sin \chi \\ 
- i e^{i \phi}\sin \gamma \sin \chi 
& \cos\gamma +i\sin\gamma\cos\chi
\end{array} 
\right), 
\end{equation*} 
where $\gamma = 2\pi -\gamma_{+}=\gamma_{-}$ has been used to simplify
the notation.  

\subsection{Cancellation of dynamical phases} 
\label{subsec:cancellation}
We closely follow Zhu and Wang's proposal\,\cite{zhu2003} 
in order to eliminate a dynamical phase. 
They proposed the use of two successive unitary operations, in which 
a dynamical phase cancels out while a geometric phase accumulates along
these two operations.  
Each unitary operation associated with a loop is characterized by the
time-dependent external field (\ref{eq:field}). 
The loop parameter corresponding to the $i$th loop is denoted by
$\bm{\Omega}_{i}(t)$ ($i=1,2$). 
Thus, in principle, we have four independent parameters in each
loop, i.e., $\omega_{i,1}$, $\omega_{i,{\rm rf}}$, $\phi_{i}$, and
$\omega_{i,0}$.  
We note that they are not always tunable in a real experimental
situation. 
\begin{figure}[tbp]
\begin{center}
\scalebox{0.5}[0.5]{\includegraphics{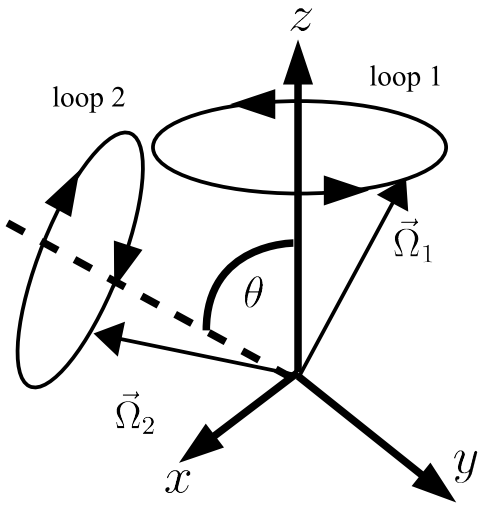}}
\end{center}
\caption{Schematic diagram of double-loop method for dynamical phase
 cancellation according to the proposal by Zhu and
 Wang\,\cite{zhu2003}. 
Two time-dependent magnetic fields are applied sequentially. 
The first magnetic field (loop 1) rotates counter-clockwise,
 while the second one (loop 2) rotates clockwise in order to eliminate a
 dynamical phase. 
\label{fig:loops}}
\end{figure}

\begin{figure}[tp]
\centering
\scalebox{0.59}[0.59]{\includegraphics{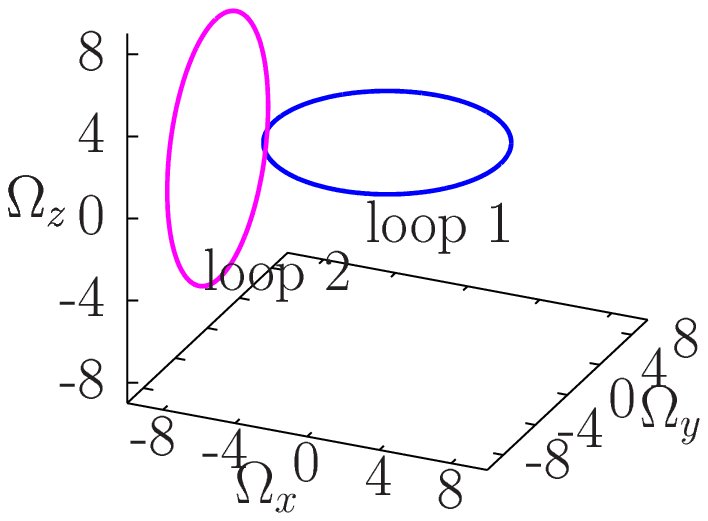}}
\scalebox{0.59}[0.59]{\includegraphics{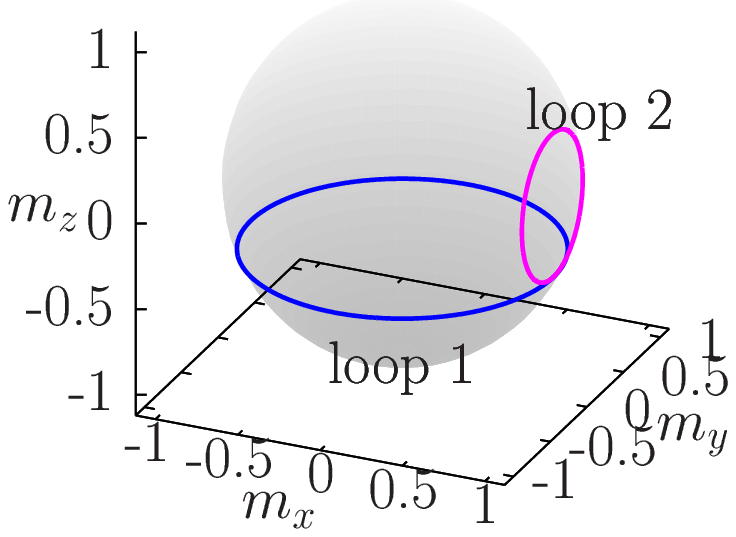}} 
\vspace{-3mm}
\flushleft
(a)\hspace{45mm}(b)
\caption{(color online) Example of Zhu-Wang's double-loop method. 
The time-dependent external fields $\bm{\Omega}_{1}(t)$ and
 $\bm{\Omega}_{2}(t)$ are shown in (a), while the closed trajectory 
on the Bloch sphere corresponding to the cyclic vectors
 $|\psi_{1,+}\rangle$ and $|\psi_{2,+}\rangle$ in (b). 
We note that these are connected and thus form one closed
 trajectory. 
We set the loop parameters $\omega_{1,1}=\omega_{2,1}=2\pi$, $\omega_{\rm rf}=0.7\times 2\pi$,
 $\omega_{1,0}=0.27\times 2\pi$, and $\omega_{2,0}=1.5\times 2\pi$ in
 Eqs.\,(\ref{loop_1}) and (\ref{loop_2}). We note that these parameters are
 calculated on the basis of a condition for nulling dynamical phases in
 Ref.\,\cite{zhu2003}. 
In this example, $\Gamma=\frac{1}{2}$ in Eq.\,(\ref{eq:ori_geop}).}
\label{fig:exampleZW}
\end{figure}

We will search for the condition under which
the dynamical phases associated with the two loops cancel each other as
shown in Fig.\,\ref{fig:loops}. 
We focus on the case in which $\phi_{i}=0$ and 
\(
\omega_{1,{\rm rf}}=\omega_{2,{\rm rf}}(\equiv \omega_{\rm rf}) >0
\) 
for simplicity. 
The first loop (loop 1) is described by
\begin{equation} 
\bm{\Omega}_1(t) 
=
\left( 
\omega_{1,1} \cos \omega_{\rm rf} t,\, 
-\omega_{1,1} \sin \omega_{\rm rf} t,\, 
\omega_{1,0} 
\right),
\label{loop_1}
\end{equation}
while the second loop (loop 2) by
\begin{equation} 
\bm{\Omega}_2(t)
= 
- 
\left( 
\omega_{2,1} \cos \omega_{\rm rf} t,\, 
-\omega_{2,1} \sin \omega_{\rm rf} t,\, 
\omega_{2,0} 
\right) \bm{R}_{y}(\theta). 
\label{loop_2}   
\end{equation} 
Let $\bm{R}_{y}(\theta) \in {\rm SO}(3)$ represent a rotation around the
$y$-axis by an angle $\theta$.  
The rotation angle $\theta$ is chosen so that the corresponding cyclic 
vectors $|{\psi_{i,\pm}}\rangle$ for these two loops satisfies 
\begin{equation}
|\psi_{1,\pm}\rangle = e^{i c}|\psi_{2,\pm}\rangle, 
\label{eq:connect_cond}
\end{equation} 
where $c \in \mathbb{R}$. 
In other words, the Bloch vectors corresponding to
$|\psi_{1,\pm}\rangle$ coincide with those to 
$|\psi_{2,\pm}\rangle$. 
Using the notation in Eq.\,(\ref{eq:nmr_rot_ham}), we find that in the
loop $i$ 
\begin{equation*} 
\bm{m}_{i} 
=
\bm{k}_{i}\bm{O}_{i},
\quad
\bm{k}_{i}
= (\sin\chi_i,0,\cos\chi_i),
\end{equation*} 
where 
\(
\bm{O}_{1} = \bm{I}_{3}
\), 
\(
\bm{O}_{2} = \bm{R}_{y}(\theta)
\), and we write 
the $3 \times 3$ unit matrix as
\(
\bm{I}_{3}
\). 
The rotation angle $\chi_{i}$ is defined as  
\begin{eqnarray}
&&
\tan \chi_1 = \frac{\omega_{1,1}}{\Delta_1}, \quad
\Delta_1 =\omega_{1,0} - \omega_{\rm rf}, 
\label{eq:def_del1_omega1}\\ 
&&
\tan \chi_2 = \frac{\omega_{2,1}}{\Delta_2}, \quad 
\Delta_2 = \omega_{2,0} + \omega_{\rm rf}. 
\label{eq:def_del2_omega2}
\end{eqnarray} 
The angle $\theta$ is explicitly given as 
\begin{equation*}
\theta = \chi_{2} - \chi_{1}.
\end{equation*} 
Figure \ref{fig:exampleZW}(a) shows an example of the time-dependent
external fields ${\bm \Omega}_{1}(t)$ and ${\bm \Omega}_{2}(t)$. 
The corresponding closed trajectory on the Bloch sphere is drawn in
Fig.\,\ref{fig:exampleZW}(b), in which the initial point corresponds to
a cyclic vector $|\psi_{1,+}\rangle$. 
Figure \ref{fig:exampleZW}(b) shows that $|\psi_{1,+}\rangle$ is not
only cyclic for loop 1 but also for the total process (i.e., loop $1$ and
loop $2$) due to the connection condition (\ref{eq:connect_cond}). 
It is necessary to search for $\omega_{i,a}$ and $ \omega_{\rm rf}$ 
($a=0,1$) so that  
\begin{eqnarray}
&&
\gamma_{\rm 1,d} +\gamma_{\rm 2,d} = 0, 
\label{eq:ori_no_dynp}\\ 
&&
\gamma_{\rm 1,g} + \gamma_{\rm 2,g} = \Gamma \pi . 
\label{eq:ori_geop}
\end{eqnarray} 
We write them more explicitly as follows: 
\begin{eqnarray} 
 \frac{\omega_{1,1}^2 + \omega_{1,0} \Delta_1}
      {\Omega_1}
&=&  
\frac{\omega_{2,1}^2 + \omega_{2,0} \Delta_2}{\Omega_2}, 
\label{eq:con_d}\\ 
\frac{\Delta_1}{\Omega_1}+\frac{\Delta_2}{\Omega_2} 
&=&2-\Gamma, 
\label{eq:con_g}
\end{eqnarray} 
where $\Omega_i = (\omega_{i,1}^2+\Delta_i^2)^{1/2}$. 
There may be many combinations of $\omega_{i,a}$ and $\omega_{\rm rf}$
for a given $\Gamma$ which satisfy the conditions (\ref{eq:con_d}) and 
(\ref{eq:con_g})\,\cite{zhu2003}. 
We note that a set of the parameters employed in
Fig.\,\ref{fig:exampleZW} is one example for the solution of
Eqs.\,(\ref{eq:con_d}) and (\ref{eq:con_g}), in which
$\Gamma=\frac{1}{2}$. 

After the elimination of a dynamical phase, we have a one-qubit
geometric quantum gate 
\begin{equation}
 V_{\rm ZW} 
= e^{i\Gamma\pi}|\psi_{1,+}\rangle\langle \psi_{1,+}|
+
e^{-i\Gamma\pi}|\psi_{1,-}\rangle\langle \psi_{1,-}|. 
\label{eq:ZW_gate}
\end{equation}

\subsection{Spin-echo approach }
\label{subsec:spin_echo}
Zhu and Wang's proposal for eliminating a dynamical phase is not
feasible for a conventional commercial NMR system where a field along
the $z$-axis is strictly constant. 
In other words, it is difficult to realize $\bm{\Omega}_2(t)$ in
Eq.\,(\ref{loop_2}). 
\begin{figure}[tp]
\begin{center}
\scalebox{0.45}[0.45]{\includegraphics{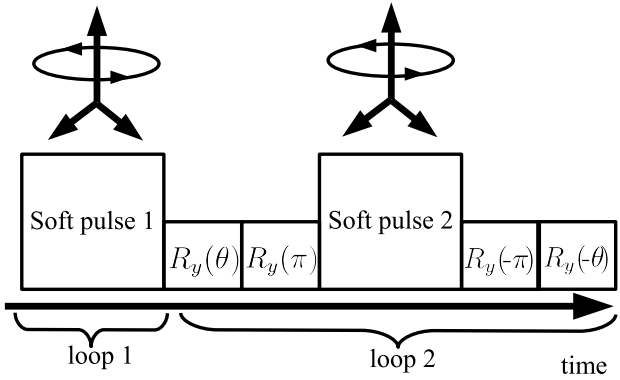}}
\end{center}
\caption{Schematic diagram of double-loop method for dynamical phase
 cancellation on the basis of a spin-echo approach.  
Two (four) soft (hard) square pulses are applied. 
We note that $R_{y}(\theta)=e^{-i\theta\sigma_{y}/2}$, in which
 $\theta=\chi_{2}-\chi_{1}$.  
\label{fig:pseqloop}}
\end{figure}
In the present paper, we propose an experimentally feasible method, in
which the loop 2 is divided into three successive steps while the loop 1
remains unchanged. 
The separation of the loop 2 is motivated by the spin-echo technique
frequently employed in NMR experiments, in which the direction of time
is reversed by an application of a pair of $\pi$-pulses. 
Three successive operations are 
(a) a rotation around the $y$-axis by 
\(
\theta (= \chi_{2}-\chi_{1})
\), 
(b) an operation corresponding to precession by a field 
\(
-\left(
\omega_{2,1} \cos \omega_{\rm rf}t,\,
\omega_{2,1} \sin \omega_{\rm rf}t,\, 
\omega_{2,0} \right)
\)
for a duration $\tau = 2\pi/|\omega_{\rm rf}|$, 
and (c) a rotation around the $y$-axis by $-\theta$.  
The rotations $R_{y}(\pm \theta)$ correspond to the basis
vector change and back as shown in Fig.\,\ref{fig:pseqloop}. 
Rotation around the $y$-axis by $\theta$ is easy to realize by
a radio-frequency (rf) pulse, which corresponds to the unitary operator
\begin{equation}
R_{y}(\theta) = e^{-i \theta \sigma_y/2}. 
\end{equation}
We emphasize here that 
\(
|\psi_{2,\pm}\rangle = R_{y}(\theta)|\psi_{1,\pm}\rangle 
\).   
We assume that the pulse duration is infinitely short for simplicity. 
It should be noted that this operation does not generate a dynamical phase
since the $y$-axis is perpendicular to both $\bm{k}_1$ and 
$\bm{k}_2$\,\cite{Ota;Kondo:2009,SuterMuellerPines1988}. 
The operation corresponds to a precession 
by a field 
\(
-\left(
\omega_{2,1}\cos\omega_{\rm rf}t,\, 
\omega_{2,1}\sin\omega_{\rm rf}t,\, 
\omega_{2,0} \right)
\)
for $\tau = 2\pi/|\omega_{\rm  rf}|$ is given as 
\begin{equation*} 
U_{2}^{\prime} =
e^{i\Omega_{2} \tau (-\bm{k}_2) \cdot\vsigma/2}, 
\end{equation*} 
from Eq.\,(\ref{eq:sol_1q}). 
The identity 
\begin{equation*}
R_{y}(-\pi) e^{i\theta \bm{k}_{2} \cdot\vsigma/2} 
R_{y}(\pi) =
e^{i\theta (-\bm{k}_{2}) \cdot\vsigma/2}, 
\end{equation*} 
implies that $U_2^{\prime}$ can be realized by a precession
under the field 
\(
 \left(
\omega_{2,1} \cos\omega_{\rm rf}t,\, 
-\omega_{2,1} \sin\omega_{\rm  rf}t,\,
\omega_{2,0} \right)
\)
for $\tau = 2\pi/|\omega_{\rm rf}|$ sandwiched by a pair of 
$\pm\pi$-pulses\,\cite{levitt}. 
We again assume that $R_{y}(\pm\pi)$ is implemented for an infinitely
short pulse for simplicity. 

Summarizing the above arguments, the total process is described by 
\(
R_{y}(-\theta)R_{y}(-\pi)U_{2}(\tau)R_{y}(\pi)R_{y}(\theta)U_{1}(\tau)
\)\,\cite{comment1}, which is equivalent to Eq.\,(\ref{eq:ZW_gate}) if
the dynamical phase is zero. 
The unitary operator $U_{i}(\tau)$ is defined as 
\(
U_{i}(\tau) 
= 
e^{i\omega_{\rm rf}\tau\sigma_{y} /2}
e^{i\Omega_{i}\bm{k}_{i}\cdot\vsigma /2}
\).  
The geometric gate which we are going to demonstrate takes the form 
\begin{equation}
U_{\rm echo} 
= R_{y}(-\pi)U_{2}(\tau)R_{y}(\pi)R_{y}(\theta)U_{1}(\tau). 
\label{eq:gate_by_pulse}
\end{equation}
Although the complete realization of Zhu and Wang's original proposal
(\ref{eq:ZW_gate}) requires $R_{y}(-\theta)$ at the end of the process
(\ref{eq:gate_by_pulse}), we can omit it for constructing a geometric quantum gate since
$R_{y}(-\theta)$ does not generate any dynamical phase here. 
We note that 
\(
V_{\rm ZW} 
= R_{y}(-\theta) U_{\rm echo}
\). 
Under the conditions (\ref{eq:ori_no_dynp}) and (\ref{eq:ori_geop}), the
matrix representation of Eq.\,(\ref{eq:gate_by_pulse}) in the
computational basis $\{|0\rangle,|1\rangle\}$ is given by  
\begin{widetext}
\begin{equation}
U_{\rm echo}
=
\left(
\begin{array}{cc}
\cos(\Gamma \pi)\cos(\theta/2)-i\sin(\Gamma \pi)\cos\Theta 
& 
-\cos(\Gamma \pi)\sin(\theta/2)-i\sin(\Gamma \pi)\sin\Theta 
\\
\cos(\Gamma \pi)\sin(\theta/2)-i\sin(\Gamma \pi)\sin\Theta 
&
\cos(\Gamma \pi)\cos(\theta/2)+i\sin(\Gamma \pi)\cos\Theta 
\end{array}
\right), 
\label{eq:mrep_gate_pulse}
\end{equation}
\end{widetext} 
where 
\(
\Theta = (\chi_{2}+\chi_{1})/2
\). 
We note that $\Gamma\pi$ is the total geometric phase. 
The pulse sequence (\ref{eq:gate_by_pulse}) leads to intuitive
understanding of the cancellation mechanism of the dynamical phase. 
Let us consider the case of $\chi_{1}=\chi_{2}$, i.e., the two loop are
completely identical, for simplicity. 
If no $\pi$-pulse is applied, the dynamical property of the loop 1 is
the same as that of the loop 2 and the total dynamical phase is the addition
between the contributions from the loops 1 and 2. 
It should be noted here that the $\pm \pi$-pulses induce the time-reversal
dynamics in the form of $U^{\prime}_{2}$ in the loop 2. 
Under the time-reversal transformation, the sign of the dynamical phase
associated with the loop 2 is inverted and hence the dynamical phase is
completely eliminated. 
It is necessary to employ different processes between the loops 1 and 2 to
prevent the cancellation of the geometric phase associated with the two loops.  
The matrix representation (\ref{eq:mrep_gate_pulse}) implies that 
$U_{\rm echo}$ contains three parameters $\Gamma$, $\theta$, and
$\Theta$. 
Due to the limitation in the control parameters, it may be
difficult to choose them independently in a standard liquid-state NMR. 
We will show that $\theta$ and $\Theta$ should be regarded as functions 
of $\Gamma$
and $\omega_{1}$ in Sec.\,\ref{subsec:implementation}, in order to
satisfy Eqs.\,(\ref{eq:ori_no_dynp}) and (\ref{eq:ori_geop}) as shown in
Table \ref{table:sol}.  
On the other hand, we are still able to use the rf phase $\phi$. 
Thus, we have the necessary number of free parameters to express
arbitrary elements of SU(2). 

\subsection{Implementation in liquid-state NMR}
\label{subsec:implementation}
We implement the double-loop scheme in liquid-state NMR.
We take different loop parameterization from that of Zhu and Wang for
ease of implementation.  
We consider the system in a rotating frame defined by $\omega_{\rm ref}$. 
Hereafter, we will denote an angular frequency $x$ measured with respect
to this rotating frame as $\tilde{x}\equiv x-\omega_{\rm ref}$. 
Thus, one can explicitly understand which quantities are controllable by
choosing a proper rotating frame. 

We take a common value 
\begin{eqnarray*}
\tilde{\omega}_{0}=\omega_{0}-\omega_{\rm ref}< 0
\end{eqnarray*}
to $\omega_{1,0}$ and $\omega_{2,0}$. 
The value of $\tilde{\omega}_{0}$ in the experiment will be explained in
Sec.\,\ref{subsec:pulse_seq}.  
Similarly, we assume that $\omega_{1,1}=\omega_{1}$ and
$\omega_{2,1}=\omega_{1}$. 
Instead of these simplification, we allow different values with respect
to $\omega_{{\rm rf}}$ between the two
loops, i.e. $\omega_{\rm rf}=\omega_{1,{\rm rf}}$ in 
Eq.\,(\ref{loop_1}) and $\omega_{\rm rf}=\omega_{2,{\rm rf}}$ in
Eq.\,(\ref{loop_2}). 
These changes do not alter
the basic strategy for searching parameters that satisfy 
$\gamma_{\rm 1,d} + \gamma_{\rm 2,d}=0$
and 
$\gamma_{\rm 1,g} + \gamma_{\rm 2,g}= \Gamma \pi$. 
We consider the two loops in the rotating frame in which the frequency 
is $\tilde{\omega}_{i,{\rm rf}}$, the amplitude $\omega_{1}$,
and the duration $\tau_i=2 \pi /|\tilde{\omega}_{i,{\rm rf}}|$, i.e.
\begin{equation*}
\tilde{{\bm \Omega}}_i(t) = 
(\omega_1 \cos \tilde{\omega}_{i, \rm rf}\,t, 
-\omega_1 \sin \tilde{\omega}_{i, \rm rf}\,t,
\tilde{\omega}_0)
\quad
(0\le t\le \tau_{i}).
\end{equation*}
The solutions $\tilde{\omega}_{i,{\rm rf}}/|\tilde{\omega}_0|$ are numerically
obtained for given $\epsilon(\equiv \omega_1/|\tilde{\omega}_0|)$ and
$\Gamma$. 

\begin{table}[bp]
\begin{tabular}{ccccccccc}
\hline \hline
$\epsilon$ 
&& $\tilde{\omega}_{1, \rm rf}/\tilde{\omega}_0$ 
&& $\tilde{\omega}_{2, \rm rf}/\tilde{\omega}_0$ 
&& $\theta$~[rad] && $\Theta $~[rad]\\
\hline \hline
0.5 && -0.6815 && 0.7803 
    && -0.7298  && -0.639 \\
0.3 && -0.8221 && 1.105  
    && -0.9571  && -0.589 \\
0.1 && -0.9422 && 1.609  
    && -1.008   && -0.542 \\ 
\hline
\end{tabular}
\caption{
\label{table:sol}
Solutions of Eqs.\,(\ref{eq:ori_no_dynp}) and (\ref{eq:ori_geop}) for
 $\Gamma = \frac{1}{2}$ and 
$\epsilon =\omega_1/|\tilde{\omega}_0| = 0.5,\,0.3,\,0.1$ in the rotating
 frame so that $\tilde{\omega}_{0}<0$. 
We note that 
$ \tau_i |\tilde{\omega}_{i, \rm rf}| =  2\pi$, 
$\theta=\chi_{2}-\chi_{1}$, 
and 
$\Theta=(\chi_{1}+\chi_{2})/2$.}
\end{table}

We summarize our parameter choice. 
First of all, we adopt a common value to $\omega_{1,0}$ and
$\omega_{2,0}$, i.e., $\tilde{\omega}_{0}$ in the rotating frame defined
as the angular frequency $\omega_{\rm ref}$. 
The value of $\omega_{1}(=\omega_{1,1}=\omega_{1,2})$ is given by 
\(
\omega_{1} = \epsilon |\tilde{\omega}_{0}|
\), in which $\epsilon$ is a positive number. 
For a given $\epsilon$ and an aimed geometric phase $\Gamma$, we can
numerically find proper $\tilde{\omega}_{i,{\rm rf}}$ so that 
\(
\gamma_{1,{\rm d}}+\gamma_{2,{\rm d}}=0
\) and 
\(
\gamma_{1,{\rm g}} + \gamma_{2,{\rm g}}=\Gamma \pi
\). 
The results for $\Gamma=\frac{1}{2}$ and $\epsilon =0.5$, $0.3$, and 0.1,
for example, are shown in Table \ref{table:sol}. 
From the observation of Eqs.\,(\ref{eq:def_del1_omega1}) and
(\ref{eq:def_del2_omega2}), the sign of $\tilde{\omega}_{1, \rm rf}$
should be opposite to the one of $\tilde{\omega}_{2, \rm rf}$. 
It should be noted that the parameters given in Table \ref{table:sol} are
compatible with this requirement. 
The resultant geometric quantum gate is
Eq.\,(\ref{eq:mrep_gate_pulse}). 
The values of $\theta(=\chi_{2}-\chi_{1})$ and
$\Theta(=(\chi_{2}+\chi_{1})/2)$ are given in Table \ref{table:sol}. 

When $\Gamma=\frac{1}{2}$, Eq.\,(\ref{eq:mrep_gate_pulse}) takes the form 
\begin{eqnarray}
U_{\rm echo}(\Theta)
&=& 
e^{-i \pi\sigma_y/2}
e^{i\Omega_2 \tau_2 \bm{k}_2 \cdot\vsigma/2} \nonumber \\
&&
\quad
\times 
e^{-i (\theta+\pi)\sigma_y/2}
e^{i\Omega_1 \tau_1 \bm{k}_1 \cdot\vsigma/2} \nonumber \\
&=& e^{-i \pi/2}
\left( \begin{array}{cr}
          \cos \Theta &  \sin \Theta \\
          \sin \Theta & -\cos \Theta
       \end{array}
\right),
\label{eq:1q_gate}
\end{eqnarray} 
which we experimentally demonstrate in the next section.

\section{Experiments}
\label{sec:experiments}
\subsection{Sample and spectrometer}
We implement a one-qubit gate described by Eq.\,(\ref{eq:1q_gate})
with a conventional commercial NMR system. 
We employed a JEOL ECA-500 NMR spectrometer\,\cite{jeol} whose 
hydrogen Larmor frequency is approximately 500\,MHz.
$^{13}$C nucleus in a 0.6\,ml, 0.2\,M sample of ${}^{13}$C-labeled 
chloroform (Cambridge Isotope) in d-6 acetone is employed 
as a qubit, while protons are decoupled by a standard decoupling 
technique, called WALTZ\,\cite{levitt}. 
We have chosen $^{13}$C-labeled chloroform for future experiments
involving two-qubit gates. 
The transverse and the longitudinal relaxation times are
$T_2 \sim 0.3\,{\rm s}$ and $ T_1 \sim 5\,{\rm s}$, respectively.
The longitudinal relaxation time is shorten by adding a small amount of 
Iron(I\!I\!I)-acetylacetonate so that a repetition rate 
can be increased. $T_2$ and $T_1$ without Iron(I\!I\!I)-acetylacetonate
are $\sim 0.3\,{\rm s}$ and $\sim 20\,{\rm s}$, respectively. 

\subsection{Pulse sequence}
\label{subsec:pulse_seq}
As we discussed in the previous section, the gate (\ref{eq:1q_gate}) 
can be realized with two rotating magnetic fields and two hard 
(short) pulses. 
The rotating fields are effectively obtained by two soft (long) pulses
which are rotating with different frequencies 
\mbox{$\tilde{\omega}_{i,\rm rf} = \omega_{i, \rm rf} - \omega_{\rm ref}$} 
($i = 1, 2$) in the rotating frame with frequency $\omega_{\rm ref}$. 
The first soft pulse (loop 1) is a usual square
pulse, while the second soft pulse (loop 2) is a
(frequency) shifted laminar square pulse (SLP)\,\cite{ECA500}. 
This SLP is employed in order to obtain the same phase $\phi$ in
Eq.\,(\ref{eq:field}) for loop 2 as that for loop 1, i.e.,
$\phi_{1}=\phi_{2}$.  

We take $|\tilde{\omega}_0| = 2 \pi \times 1000\,{\rm rad}/{\rm s}$
and $\phi_{i} = 0$ throughout the experiments. 
The condition $\phi_{i} = 0$ is taken for 
simplicity as mentioned in the beginning of 
Sec.\,\ref{subsec:cancellation}.
We independently calibrate the strengths of the soft and
hard pulses in order to minimize a non-linearity error in setting the
rf pulse amplitude. 
The duration $t_{\rm hp}$ of a hard $ \pi $-pulse is set to 
$ 21.6\,\mu{\rm s}$ throughout the experiments. 
We ignore $t_{\rm hp}$ in setting the phase of the second soft pulse, 
which is justified by the fact that 
$t_{\rm hp} |\tilde{\omega}_0| \ll 2\pi$. 
The precision of pulse duration control is 100\,ns. 
The durations $t_{i, \rm sp}$ of two soft pulses are set to
\begin{eqnarray*}
t_{i, \rm sp} |\tilde{\omega}_{i, \rm rf}|= 2 \pi. 
\end{eqnarray*}
We demonstrate three different gates with $\epsilon=0.5,0.3$ and $0.1$. 
We note that the phase of the second hard pulse 
corresponding to $R_{y}(-\pi)$ must be adjusted, 
presumably because the oscillator in the NMR spectrometer is 
disturbed in generating a SLP. 
It should be recalled that a SLP employs intensive phase modulation. 

\subsection{Results}
Implemented gates with $\Gamma = \frac{1}{2}$ are evaluated by performing  
quantum process tomography\,\cite{qpt}. 
The practical details are explained in Ref.\,\cite{kondo}. 
A quantum process ${\mathcal E}$, such as a gate operation or relaxation
process, is 
\begin{equation*}
\rho \mapsto 
{\mathcal E}(\rho) 
= \frac{\sum_{k} E_{k} \rho E_{k}^\dagger}{\sum_{k}E^{\dagger}_{k}E_{k}}
\end{equation*}  
in the operator sum (or Kraus) representation\,\cite{Kraus:1983,op_sum_r}. 
When all $E_{k}$'s are determined, $\mathcal{E}$ is considered 
to be identified. 
This identification is called quantum process tomography. 

\begin{figure}[tbp]
\centering
\scalebox{0.21}[0.21]{\includegraphics{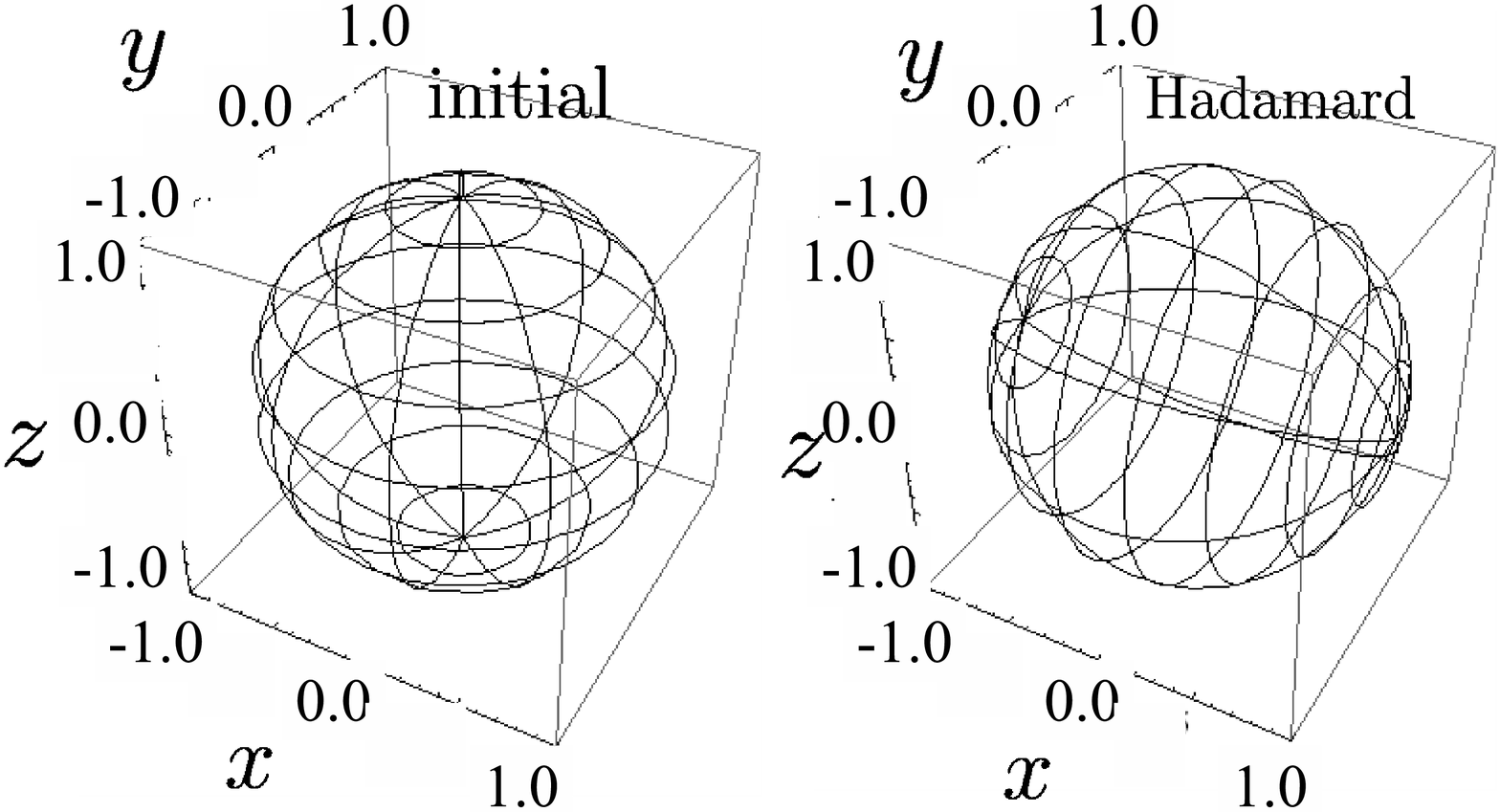}}
\vspace{-7mm}
\begin{flushleft}
(a)
\end{flushleft} 
\scalebox{0.21}[0.21]{\includegraphics{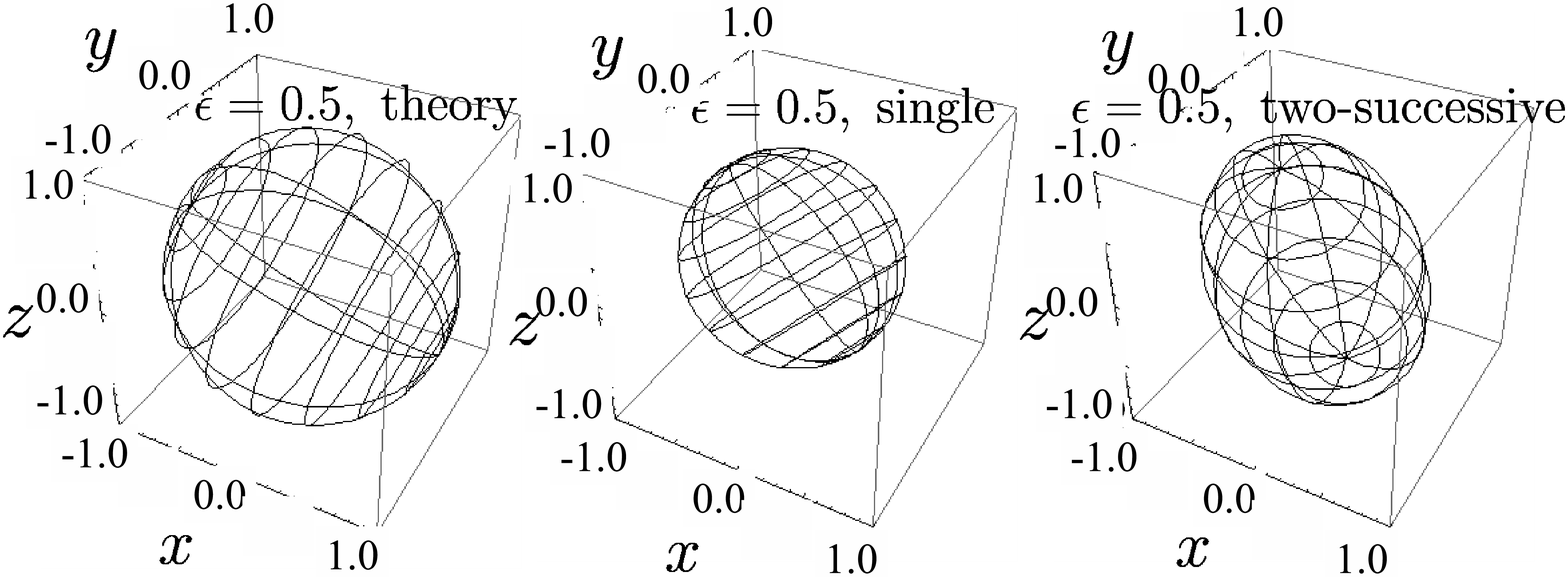}}
\vspace{-7mm}
\begin{flushleft}
(b)
\end{flushleft} 
\scalebox{0.21}[0.21]{\includegraphics{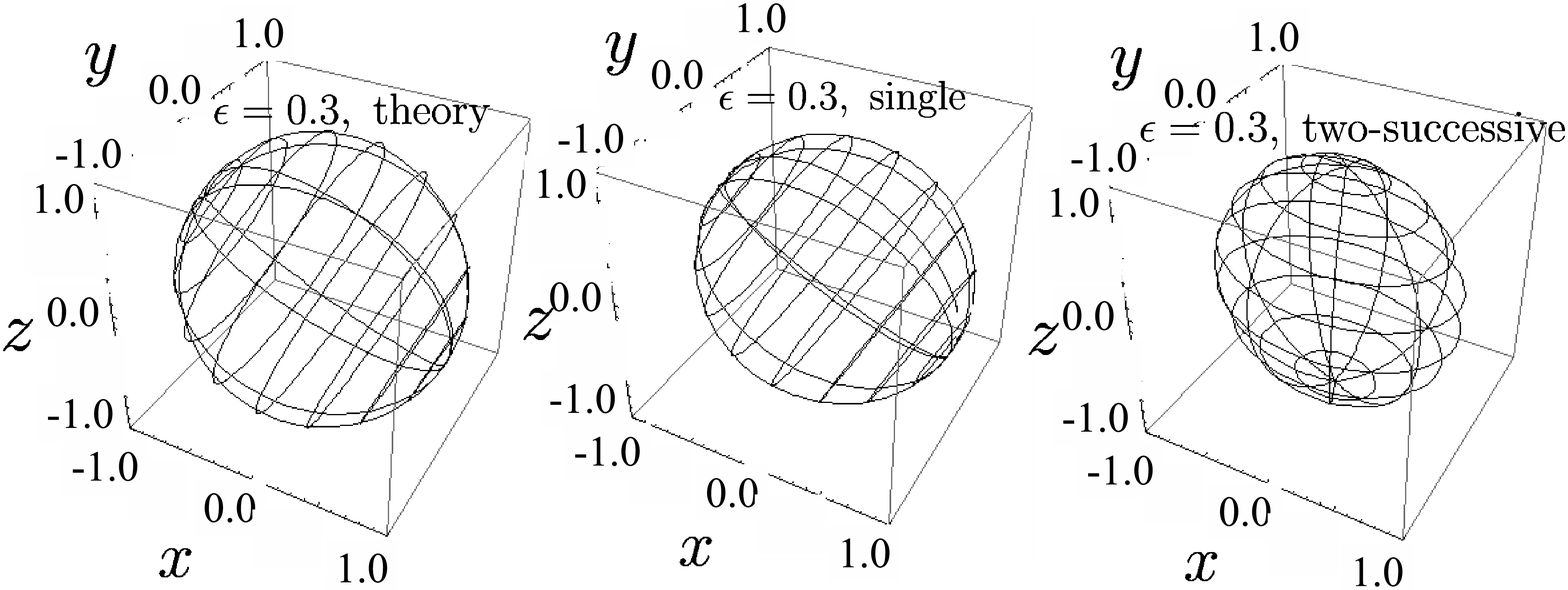}}
\vspace{-7mm}
\begin{flushleft}
(c)
\end{flushleft} 
\scalebox{0.21}[0.21]{\includegraphics{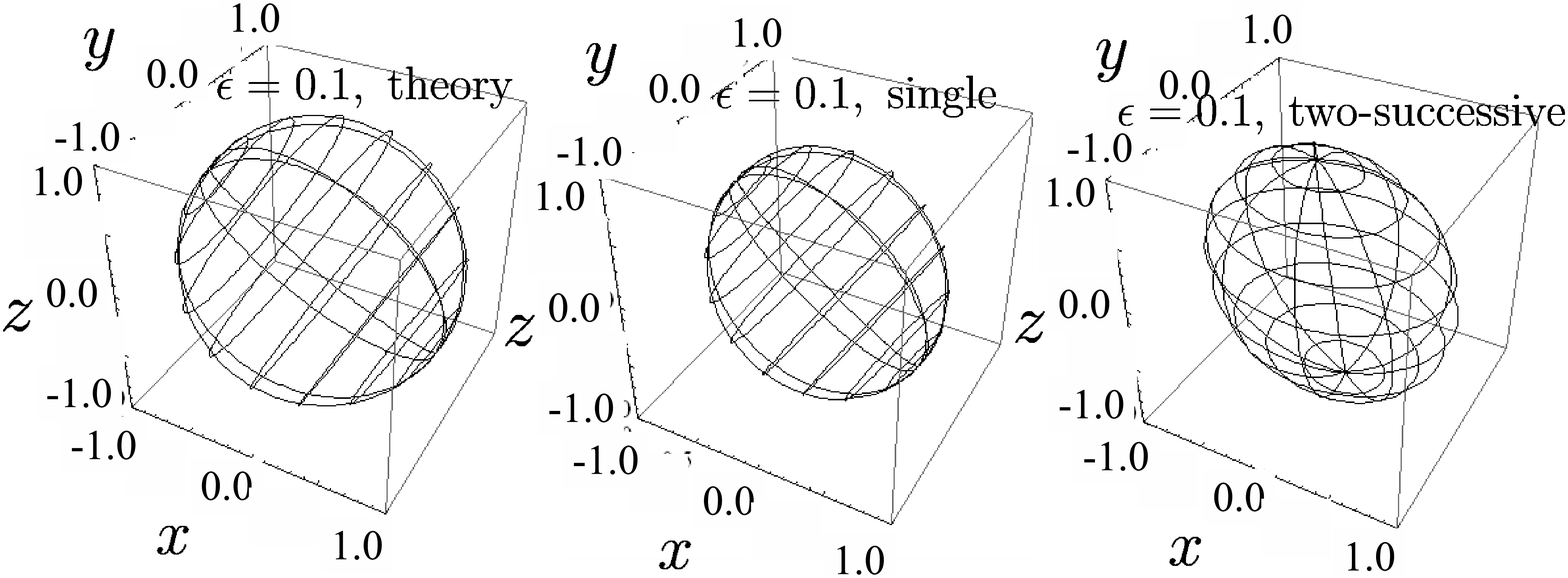}}
\vspace{-7mm}
\begin{flushleft}
(d)
\end{flushleft} 
\caption{
\label{fig:results}
Gate operations visualized. The Bloch sphere 
in (a) is mapped to the surfaces in (b), (c), and (d) under the gates with
 $\epsilon = 0.5, 0.3$ and $0.1$, respectively. 
The right surface in (a) is an expected Bloch sphere when 
$\Theta =-\pi/4$, which corresponds to the Hadamard gate. 
Each left panel in (b), (c), and (d) corresponds to the theoretical
 final state. 
The middle panels are the results for the single gate operation 
$U_{\rm echo}(\Theta)$. 
The right panels are for the two-successive (double) gate
 operations. }
\end{figure}

\begin{table}[tbp]
\begin{tabular}{c|cc|cccc}
\hline \hline
  $\epsilon$ 
& $F_{\rm e}(I_{0}, {\mathcal E}\circ\mathcal{U}_{\rm echo}^{-1})$ 
& ${\rm Tr}[{\mathcal E}(I_{0})] $  &
& $F_{\rm e}(I_{0}, {\mathcal E}^2)$ 
& ${\rm Tr}[{\mathcal E}^2(I_{0})] $ \\
\hline \hline
0.5 & 0.75 & 1.00 && 0.74 &  1.02 \\
0.3 & 0.88 & 1.08 && 0.83 &  1.07 \\
0.1 & 0.84 & 1.07 && 0.85 &  1.06 \\\hline
\end{tabular}
\caption{
\label{tab:entangle}
The entanglement fidelities for single and double operations 
with $\epsilon = 0.5$, $0.3$, and $0.1$. }
\end{table}
The Bloch sphere in Fig.\,\ref{fig:results}(a) is mapped under the gate
operations to the surfaces in Figs.~\ref{fig:results}(b), (c) and
(d), which correspond to $\epsilon = 0.5, 0.3$ and $0.1$, respectively. 
If the gate operations are perfect, the surfaces are the spheres of unit
radius (i.e., the Bloch sphere). 
The left panel of each row shows the theoretical final state. 
In the middle panels, the results for the single gate operation are
shown. 
Finally, the right panels are for the two-successive (double) gate
operation. 
The Hadamard gate obtained when $\Theta = -\pi/4$ is, for comparison, shown 
in the right pannel of Fig.\,\ref{fig:results}(a). 
From these figures, we find that $U_{\rm echo}(\Theta)$ 
in Eq.\,(\ref{eq:1q_gate}) is implemented although it is not perfect.  

We numerically evaluated the fidelity of the implemented gate using the
entanglement fidelity\,\cite{kondo,op_sum_r} given by 
\begin{eqnarray*}
F_{\rm e}(I_{0},\mathcal{E}\circ\mathcal{U}_{\rm echo}^{-1})
&=&
\frac{\sum_{k}|{\rm Tr}[E_{k}U_{\rm echo}(-\Theta)I_{0}]|^2}
{{\rm Tr}[\sum_k E_k U_{\rm echo}
(-\Theta)I_0 U_{\rm echo}^\dagger(-\Theta) E_k^\dagger ]} \nonumber \\
&=&
\frac{\sum_{k}|{\rm Tr}[E_{k}U_{\rm echo}(-\Theta)I_{0}]|^2}
{{\rm Tr}(\sum_k E_k I_0 E_k^\dagger )},
\end{eqnarray*}
where $\mathcal{U}_{\rm echo}$ is a super operator corresponding to the
unitary operator $U_{\rm echo}(\Theta)$ (i.e., 
\(
\mathcal{U}_{\rm echo}(\rho) = U_{\rm echo}(\Theta)\rho U^{\dagger}_{\rm echo}(\Theta)
\)),  
$I_{0} = \openone/2$, and $\openone$ is the identity matrix of dimension $2$. 
One can find that 
\(
F_{\rm e}(I_{0}, {\mathcal E}\circ\mathcal{U}_{\rm echo}^{-1}) = 1
\)
when the gate operation $\mathcal{E}$ is perfect.  
In the case of two successive gate 
operation, $F_{\rm e}(I_{0}, {\mathcal E}^2)$ gives a measure of 
the fidelity since $[U_{\rm echo}(\Theta)]^{2} = -\openone$. 
The entanglement fidelities corresponding to the gate operations 
are summarized in Table \ref{tab:entangle}. 

The fidelities of the demonstrated gates are not high. 
This may be attributed to the inhomogeneous rf field. 
The free induction decay signal of the thermal state 
after a $5\pi/2$-pulse, which corresponds to the operation 
\(
e^{-5\pi\sigma_{x}/4}
\) for example, reduces to about $85$\% of that after 
a $\pi/2$-pulse, which corresponds to the operation 
\(
e^{-\pi\sigma_{x}/4} 
\) for example. 
This fact indicates that there is some
rf field inhomogeneity which may account for most of the reduction in 
the fidelities in Table \ref{tab:entangle}. 
Pulse sequences in usual NMR operations are designed so that the 
rf field inhomogeneity does not affect measurements, for example,
by employing composite pulses. Such techniques are not available 
in our experiments. 

\section{Summary}
\label{sec:summary}
We demonstrated the elimination of the dynamical phase and
the implementation of the quantum gates with pure nonadiabatic
geometric phases in a liquid-state NMR quantum computer, based on the
double-loop method.  
By means of a spin echo technique, we modified the original proposal so
that quantum gates are implemented in a standard high precession NMR
system for chemical analysis. 
We have proposed and experimentally verified an alternative method to eliminate
dynamical phase. 
The extension of the present method to two-qubit operations is an
important future work\,\cite{comment2}. 
We believe that our work is the first step toward physical realization of
working geometric quantum gates and further efforts should be made for
improvement of the gates. 

\begin{acknowledgments} 
This work was supported by ``Open Research Center'' Project for 
 Private Universities: Matching fund subsidy from MEXT (Ministry of 
 Education, Culture, Sports, Science and Technology).  
MN's work is supported in part by Grant-in-Aid for Scientific Research
 (C) from JSPS (Grant No. 19540422).  
\end{acknowledgments}


\begin{thebibliography}{99}
\bibitem{comment1}
When the rf phases for the loop 1 and 2 are not zero, instead of 
\(
 R_{y}(\theta)
\), one has to perform 
\(
e^{-i\phi_{2}\sigma_{z}/2}R_{y}(\theta)e^{i\phi_{1}\sigma_{z}}
\), where $\phi_{i}$ is the rf phase for the loop $i$. 
In addition, the $\pi$-pulse $R_{y}(\pi)$ has to be also replaced with 
\(
e^{-i\phi_{2}\sigma_{z}/2}R_{y}(\pi)e^{i\phi_{2}\sigma_{z}/2}
\).
\end{thebibliography}
\end{document}